\documentclass[12pt, draftclsnofoot, onecolumn]{IEEEtran}

\usepackage{cite}
\usepackage{hyphenat}
\usepackage{graphicx}
\usepackage[T1]{fontenc}
\usepackage[latin1]{inputenc}
\usepackage{amsmath,amsfonts,amsbsy,amssymb}
\usepackage{mathrsfs}
\usepackage[ruled]{algorithm}
\usepackage{algorithmic}
\usepackage[nolist]{acronym}
\usepackage{float}
\usepackage{tipa}
\usepackage{dsfont}
\usepackage{tabularx}

\usepackage[absolute,overlay]{textpos}
\usepackage{tikz}
\usetikzlibrary{topaths}
\usetikzlibrary{calc}
\usetikzlibrary{shapes,trees,decorations}
\usetikzlibrary{arrows}
\usetikzlibrary[shadows]
\usetikzlibrary{positioning}
\usetikzlibrary{matrix}
\definecolor{darkblue}{rgb}{.1,.1,.6}

\newtheorem{definition}{Definition}
\newtheorem{proposition}{Proposition}

\newtheorem{remark}{Remark}

\newcommand{\hlchange}[1] {\textcolor{black}{#1}}

\begin{document}

% Paper title.
\title{Effects of Relay Selection Strategies on the Spectral Efficiency of
Wireless Systems with Half- and Full-duplex Nodes}
%
%\title{Contention-based Geographic Forwarding Strategies for Wireless Sensors Networks}
%
\author{\IEEEauthorblockN{Carlos H.~M. de~Lima~\IEEEmembership{Member, IEEE},
Hirley Alves~\IEEEmembership{Member, IEEE}, Pedro. H. J. Nardelli, and Matti
Latva-aho~\IEEEmembership{Senior Member, IEEE}} 
    
\thanks{C. H. M. de Lima is with S\~ao Paulo State University (UNESP), S\~ao
    Jo\~ao da Boa Vista, Brazil. Email: carlos.lima@sjbv.unesp.br.
    
    H. Alves, P. H. J. Nardelli and M. Latva-aho are with the Centre for Wireless
    Communications (CWC), University of Oulu, Finland.  E-mail:firstname.lastname@oulu.fi. 
    
    The research leading to these results has received funding from Strategic Research Council/Aka BCDC Energy (Grant n.292854), Aka/SAFE (Grant n.303532) and CNPq/Universal program (Brazil). 
    }
%\thanks{Copyright (c) 2012 IEEE. Personal use of this material is permitted.
%However, permission to use this material for any other purposes must be
%obtained from the IEEE by sending a request to pubs-permissions@ieee.org}
}

% Make the title area.
\maketitle

\begin{acronym}[mmmmm]
	% \acro{IP}{Internet Protocol}
	\acro{3GPP}[$3$GPP]{$3^\mathrm{rd}$ Generation Partnership Project}
	\acro{ABS}{Almost Blank Sub-frame}
	\acro{ADSL}{Asymmetric Digital Subscriber Line}
	\acro{ALBA-R}{Adaptive Load-Balanced Algorithm Rainbow}
	\acro{ALBA}{Adaptive Load-Balanced Algorithm}
	\acro{ALOHA}[ALOHA]{}
	\acro{APDL}{Average Packet Delivery Latency}
	\acro{AP}{Access Point}
	\acro{ASE}{Average Spectral Efficiency}
	\acro{BAP}{Blocked Access Protocol}
	\acro{BB}{Busy Burst}
	\acro{BC}{Broadcast Channel}
	\acro{BPP}{Binomial Point Process}
	\acro{BS}{Base Station}
	\acro{CAA}{Channel Access Algorithm}
	\acro{CAPEX}{Capital Expenditure}
	\acro{CAP}{Channel Access Protocol}
	\acro{CCDF}{Complementary Cumulative Distribution Function}
	\acro{CCI}{Co-Channel Interference}
	\acro{CDF}{Cumulative Distribution Function}
	\acro{CDMA}{Code Division Multiple Access}
	\acro{CDR}{Convex Lenses Decision Region}
	\acro{CF}{Characteristic Function}
	\acro{CGF}{Contention-based Geographic Forwarding}
	\acro{CM}{Coordination Mechanism}
	\acro{COMP}{Coordinated Multi-Point}
	\acro{CPICH}{Common Pilot Channel}
	\acro{CRA}{Conflict Resolution Algorithm}
	\acro{CRD}{Contention Resolution Delay}
	\acro{CRI}{Contention Resolution Interval}
	\acro{CRP}{Contention Resolution Protocol}
	\acro{CRS}{Cell-specific Reference Signal}
	\acro{CR}{Contention Resolution}
	\acro{CSMA/CA}{Carrier Sense Multiple Access with Collision Avoidance}
	\acro{CSMA/CD}{Carrier Sense Multiple Access with Collision Detection}
	\acro{CSMA}{Carrier Sense Multiple Access}
	\acro{CTM}{Capetanakis-Tsybakov-Mikhailov}
	\acro{CTS}{Clear To Send}
	\acro{D2D}{Device-To-Device}
	\acro{DAS}{Distributed Antenna System}
	\acro{DCF}{Distributed Coordination Function}
	\acro{DER}{Dynamic Exclusion Region}
	\acro{DHCP}{Dynamic Host Configuration Protoco}
	\acro{DL}{Downlink}
	\acro{DSL}{Digital Subscriber Line}
	\acro{E2E}{End-to-End}
	\acro{EDM}{Euclidean Distance Matrix}
	\acro{ES}{Evaluation Scenario}
	\acro{FAP}{Femto Access Point}
	\acro{FBS}{Femto Base Station}
	\acro{FD}{Full-Duplex}
	\acro{FDD}{Frequency Division Duplexing}
	\acro{FDM}{Frequency Division Multiplexing}
	\acro{FDR}{Forwarding Decision Region}
	\acro{FFR}{Fractional Frequency Reuse}
	\acro{FG}{Frequency Group}
	\acro{FPP}{First Passage Percolation}
	\acro{FUE}{Femto User Equipment}
	\acro{FU}{Femtocell User}
	\acro{GF}{Geographic Forwarding}
	\acro{GLIDER}{Gradient Landmark-Based Distributed Routing}
	\acro{GPSR}{Greedy Perimeter Stateless Routing}
	\acro{GeRaF}{Geographic Random Forwarding}
	\acro{HC}{Hard Core}
	\acro{HD}{Half-Duplex}
	\acro{HDR}[HDR]{High Data Rate}
	\acro{HII}{High Interference Indicator}
	\acro{HNB}{Home Node B}
	\acro{HN}[HetNet]{Heteronegeous Network}
	\acro{HOS}{Higher Order Statistics}
	\acro{HUE}{Home User Equipment}
	\acro{ICIC}{Inter-Cell Interference Coordination}
	\acro{IEEE}[IEEE]{}
	\acro{IMT}{International Mobile Telecommunications}
	\acro{IP}{Interference Profile}
	\acro{ITU}{International Telecommunication Union}
	\acro{KPI}{Key Performance Indicators}
	\acro{LN}{Log-Normal}
	\acro{LTE}{Long Term Evolution}
	\acro{LoS}{Line-of-Sight}
	\acro{M2M}{Machine-to-Machine}
	\acro{MACA}{Multiple Access with Collision Avoidance}
	\acro{MAC}{Medium Access Control}
	\acro{MBS}{Macro Base Station}
	\acro{MT}{Mellin Transform}
	\acro{MTC}{Machine Type Communication}
	\acro{MGF}{Moment Generating Function}
	\acro{MIMO}{Multiple-Input Multiple-Output}
	\acro{MPP}{Marked Point Process}
	\acroplural{MPP}[MPPs]{Marked Point Processes}
	\acro{MRC}{Maximum Ratio Combining}
	\acro{MS}{Mobile Station}
	\acro{MUE}{Macro User Equipment}
	\acro{MU}{Macrocell User}
	\acro{NB}{Node B}
	\acro{NLoS}{Non Line-of-Sight}
	\acro{NRT}{Non Real Time}
	\acro{OFDMA}{Orthogonal Frequency Division Multiple Access}
	\acro{OOP}{Object Oriented Programming}
	\acro{OPEX}{Operating Expenditure}
	\acro{OP}{Outage Probability}
	\acro{OS}{Order Statistic}
	\acro{PBS}{Pico Base Station}
	\acro{PCI}{Physical Cell Indicator}
	\acro{PC}{Power Control}
	\acro{PDF}{Probability Density Function}
	\acro{PDSR}{Packet Delivery Success Ratio}
	\acro{PGF}{Probability Generating Function}
	\acro{PMF}{Probability Mass Function}
	\acro{PPP}{Poisson Point Process}
	\acroplural{PPP}[PPPs]{Poisson Point Processes}
	\acro{PP}{Point Process}
	\acro{PRM}{Poisson Random Measure}
	\acro{PSS}{Primary Synchronization Channel}
	\acro{QoS}{Quality of Service}
	\acro{RAS}{Random Access System}
	\acro{RAT}{Radio Access Technology}
	\acro{RA}{Random Access}
	\acro{RCA}{Random Channel Access}
	\acro{RD}[R$\&$D]{Research $\&$ Development}
	\acro{REB}{Range Expansion Bias}
	\acro{RE}{Range Expansion}
	\acro{RF}{Radio Frequency}
	\acro{RIBF}{Regularized Incomplete Beta Function}
	\acro{RMA}{Random Multiple-Access}
	\acro{RNTP}{Relative Narrowband Transmit Power}
	\acro{RN}{Relay Node}
	\acro{RRM}{Radio Resource Management}
	\acro{RSA}{Relay Selection Algorithm}
	\acro{RSRP}{Reference Signal Received Power}
	\acro{RSSI}{Received Signal Strength Indicator}
	\acro{RSS}{Received Signal Strength}
	\acro{RS}{Relay Selection}
	\acro{RTS}{Request to Send}
	\acro{RT}{Real Time}
	\acro{RV}{Random Variable}
	\acro{SC}{Selection Combining}
	\acro{SDR}{Sectoral Decision Region}
	\acro{SF}{Sub-Frame}
	\acro{SG}{Stochastic Geometry}
	\acro{SI}{Self-Interference}
	\acro{SIC}{Successive Interference Cancellation}
	\acro{SINR}{Signal-to-Interference plus Noise Ratio}
	\acro{SIR}{Signal-to-Interference Ratio}
	\acro{SLN}{Shifted Log-Normal}
	\acro{SMP}{Semi-Markov Process}
	\acroplural{SMP}[SMPs]{Semi-Markov Processes}	
	\acro{SM}{State Machine}
	\acro{SNR}{Signal to Noise Ratio}
	\acro{SON}{Self-Organizing Network}
	\acro{SPP}{Spatial Poisson Process}
	\acroplural{SPP}[SPPs]{Spatial Poisson Processes}
	\acro{SSS}{Secondary Synchronization Channel}
	\acro{STA}{Splitting Tree Algorithm}
	\acro{TAS}{Transmit Antenna Selection}
	\acro{TCP}{Transmission Control Protocol}
	\acro{TC}{Transmission Capacity}
	\acro{TDD}{Time Division Duplexing}	
	\acro{TDMA}{Time Division Multiple Access}	
	\acro{TS}{Terminal Station}
	\acro{TTI}{Transmission Time Interval}
	\acro{TTT}{Time To Trigger}
	\acro{UDM}{Unit Disk Model}
	\acro{UD}{Unit Disk}
	\acro{UE}{User Equipment}
	\acro{ULUTRANSIM}[UL UTRANSim]{R6 Uplink UTRAN Simulator}
	\acro{UL}{Uplink}
	\acro{UML}{Unified Modeling Language}
	\acro{UoI}{User of Interest}
	\acro{UMTS}{Universal Mobile Telecommunications System}
	\acro{WCDMA}{Wideband Code Division Multiple Access}
	\acro{WSN}{Wireless Sensor Network}
	\acro{iid}[\textup{i.i.d.}]{independent and identically distributed}
\end{acronym}

%%\vspace{-1cm}
\begin{abstract}
    This work proposes an analytical framework to study how relay selection
    strategies perform in half- and full-duplex deployments by combining
    renewal theory and stochastic geometry. 
    Specifically, we assume that the network nodes -- operating in either half-
    or full-duplex mode --  are scattered according to a two-dimensional
    homogeneous Poisson point process to compute the relay selection cost by
    using a semi-Markov process.
    Our results show: ($i$) fixed relay outperforms the reactive option in
    either cases, ($ii$) the performance of both reactive and fixed relay
    strategies depends on the self-interference attenuation in full-duplex
    scenarios, evincing when they outperform the half-duplex option, and
    ($iii$) the reactive relay selection suffers from selecting relays at hop
    basis, while the fixed relay selection benefits most from the full-duplex
    communication.
\end{abstract}
%%\vspace{-0.5cm}
%% \begin{keywords}
%%     %
%%     Full-duplex, relay selection, renewal theory, stochastic geometry
%% %
%% \end{keywords}
%\keywords{Full-duplex, relay selection, renewal theory, stochastic geometry}
\acresetall

\section{Introduction}
\label{SEC:INTRODUCTION}

In the recent years, \ac{FD} communication has gained considerable attention
from both academy and industry \cite{ART:SAVHARWAL-JSAC14,ART:LIU-CSTO15}.
By allowing simultaneous transmission and reception on the same frequency band,
\ac{FD} networks can potentially double the spectral efficiency compared to
current \ac{HD} schemes.
However, such potential is harmed by \ac{SI} from the transmit to receive
antenna \cite{ART:SAVHARWAL-JSAC14,ART:LIU-CSTO15}.
Due to recent advances in antenna design associated with analog and digital
interference cancellation, most of the \ac{SI} can be mitigated, and therefore
\ac{FD} communication becomes feasible and are a step towards meeting high
demands of spectral efficiency of 5G \cite{KattiCM2014}. 

Current works have shown that \ac{FD} is a viable solution for small cell
deployments \cite{ART:LIU-CSTO15}, due to not only cost and size constraints
but also to compatibility with current \ac{HD} systems.
Then, due to the \ac{FD} capability, the {base station} can, for instance,
simultaneously schedule uplink and downlink transmissions, see for instance
\cite{PROC:LIMA-VTC2015}. 
Moreover, {base station} can also act as relays for the legacy network as well
as for its own subscribed users.
Cooperative diversity appeared as a way to combat the effects of the fading, by
allowing a single antenna user to experience spatial diversity
\cite{GomezCST2012}.
Such schemes perform even better when the relay operates in a \ac{FD} fashion
\cite{AlvesWCL2013}. 

%\normalmarginpar
%\marginnote{R$1$C$1$}[-2ex]
Bearing this in mind, and in the context of a dense deployment of small cells,
relaying selection become an attractive solution given a trade-off between
complexity and efficiency \cite{ART:LIU-CSTO15, BletsasJSAC2006}.
Relay selection was initially proposed in \cite{BletsasJSAC2006} for \ac{HD}
cooperative networks and has gained considerable attention since. 
\hlchange{Nonetheless, most of the works so far have focused on HD selection algorithms,
except for \cite{KrikidisTWC2012, RuiHouZhouEL2010}
where the authors focus on the performance of a three-node relaying scenario.
For instance, \cite{KrikidisTWC2012} focus on selection algorithms for \ac{FD}
amplify-and-forward and model the residual \ac{SI} at relay as a Rayleigh
random variable.}
While in \cite{RuiHouZhouEL2010}, the authors address the performance of a FD
decode-and-forward protocol and assume simpler residual \ac{SI} model (constant
attenuation factor). 

%\normalmarginpar
%\marginnote{R$3$C$1$}[-2ex]
\hlchange{In this work, we assess the performance at the network level of a
relay selection procedure for cooperative diversity protocols in FD mode.
To do so, we model the relaying selection procedure as a semi-Markov process
and then we investigate the impact of the \ac{SI} and \ac{CCI} on the network
performance.
Moreover, a dynamic relay selection procedure is considered where a suitable
relay is selected at each hop and the cost of this selection procedure is
incorporated into the achievable rate.
And yet, we resort to stochastic geometry to model network deployments and
capture the dynamics of the network, more specifically the aggregate
interference at the receiver, which can be characterized through a
cumulant-based framework introduced in
\cite{ART:LIMA-TWC12A,ART:LIMA-TVT12B}.}

\hlchange{Our main contributions are summarized as follows:
\begin{itemize}
    \item The standard binary tree algorithm is used to implement a relay
        selection procedure for \ac{FD} networks; 
    \item The framework originally introduced in \cite{PROC:MARCHENKO-VTC11} is
        extended to account for the \ac{FD} relaying operation;
    \item The performance of the aforesaid relay selection protocol are
        evaluated and compared to its \ac{HD} counterpart;
    %
    %% \reversemarginpar
    %% \marginnote{R$3$C$5$}[-2ex]
    \item {Different from \cite{KrikidisTWC2012, RuiHouZhouEL2010}, our network
        level analysis accounts for a more detailed model of the residual
        \ac{SI} based on \cite{PROC:DUARTE-ASILOMAR2010} and also characterizes
        the \ac{CCI} at the receiver.
        Actually, instead of modeling the residual SI by a constant value, we
        consider a more detailed model through a random variate drew from a
        Ricean distribution with large $K$-factor.}
\end{itemize}}

%% \reversemarginpar
%% \marginnote{R$1$C$6$}[-2.25cm]
The remainder of this paper is organized as follows:
Section~\ref{SEC:SYSTEM_MODEL} describes the problem under investigation and
introduces the system model.
Section~\ref{SEC:SG_MODEL} describes the network topology and introduces our
analytical framework based on the stochastic geometry, while
Section~\ref{SEC:RSA} characterizes the relay selection procedures and the
probability generating function concept.
Section \ref{SEC:SMP} introduces the semi-Markov process that is used to
evaluate the performance of the proposed relay selection mechanism.
In Section~\ref{SEC:NUMERICAL_RESULTS}, performance metrics are introduced and
then utilized to obtain the numerical results and carry out their discussion.
We draw conclusions and final remarks in Section~\ref{SEC:CONCLUSION}.

%% \reversemarginpar
%% \marginnote{R$1$C$6$ R$2$C$1$}[0cm]

\section{{Problem Description and System model}}
\label{SEC:SYSTEM_MODEL}

As aforesaid, full duplex networks constitute a promising alternative to
achieve high data rates in the upcoming $5$G systems.
In such networks, cooperative nodes benefit most from the inherent rate
increase allowed by transmitting and receiving simultaneously.
{Typically, available literature concentrates on assessing the
throughput that is attainable by combining full duplex and cooperative
diversity protocols, and overlooks the impact of the underlying relay selection
on the final achievable data rate \cite{KrikidisTWC2012}.}

%% \reversemarginpar
%% \marginnote{R$1$C$1$}[-2ex]
{In this paper, we contribute to previous results by introducing an
    analytical framework that combines stochastic geometry and semi-Markov
    processes to account for the cost of selecting a suitable relay.
On that regard, we initially use stochastic geometry to derive the
distributions of the received power and aggregate interference experienced by
the receiver of interest in Section \ref{SEC:SG_MODEL}.
Thereafter, we use the \ac{PGF} concept \cite{PROC:LIMA-ITW09, ART:LIMA-SJ2016}
to characterize the \ac{CRI} (holding time) of the {relay selection algorithms}
as can be seen in Sections \ref{SEC:RSA} and \ref{SEC:SMP}.
Our investigations extend \cite{PROC:MARCHENKO-VTC11} in which Authors
introduce an analytic framework based on semi-Markov processes with constant
holding time and apply it to study an one-dimensional network model.}

%% \newpage
We assess the performance of interference networks operating in \ac{FD} mode
under composite fading channel -- Lognormal shadowing and Nakagami-$m$ fading.
The \ac{DL} of a traditional \ac{HD} network constitutes our benchmark scenario
wherein the user of interest is interfered by surrounding small cells.
Base stations independently schedule a random user terminal in every
transmission interval.
All communicating nodes use antennas with omni-directional radiation pattern.
Base stations and \acp{UE} are also assumed to have full buffer and symmetric
traffic patterns \cite{BOOK:BERTSEKAS-PRENTICE92}.

\begin{figure}[t!]
    \centering
    \includegraphics[width=.80\columnwidth]{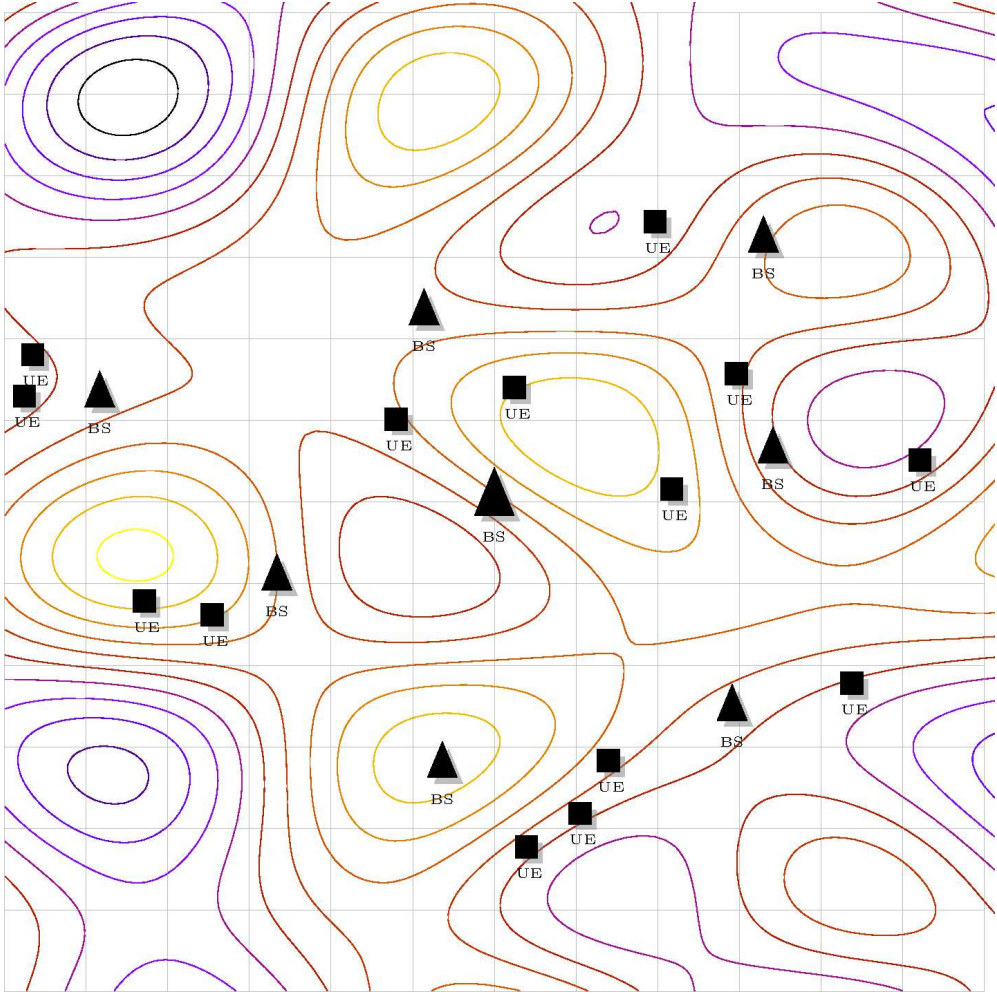}
    %\vspace{-2mm}
    \caption{Illustration of a random (bi-dimensional) deployment of base
        stations and user terminals over an arbitrary network area.
        Shaded squares represent user terminals, while shaded up-triangles
        depict base stations.
        A heat map represents the corresponding random composite fading channel
    where the fading intensity varies from red/strong to dark blue/weak.}
    \label{FIG:HN_GRID}
\end{figure}

Consider all channels are quasi-static and undergo a composite fading
distribution \cite{BOOK:STUBER-SPRINGER11}, which is approximated by a
{lognormal} distribution with mean and variance (in logarithmic scale) given by
$\mu_\mathrm{dB} = \xi \left [ \psi\left ( m \right ) - \ln\left ( m
\right)\right ] + \mu_{\Omega_p}$ and $\sigma_\mathrm{dB}^2 = \xi^{2}\zeta\left
( 2, m \right) + \sigma_{\Omega_p}^2$, where $\psi\left ( m \right )$ is the
Euler psi function and $\zeta\left ( 2, m \right )$ is the generalized Riemann
zeta function \cite{BOOK:ABRAMOWITZ-DOVER03}.
Radio links are affected by path-loss, large-scale shadowing and multi-path
fading which are assumed to be mutually independent and multiplicative
phenomena \cite{ART:ILOW-TSP98}.
The received power at the user of interest $u_0$ from an arbitrary transmitter
$b_i$ located $r_{i0}$ meters away is
%
%\begin{align}
%    %
    $Y_{i0} = p_{i0}\, \thinspace r_{i0}^{-\alpha} x_{i0}$,
%    %
%    \label{EQ:INTERFERENCE_COMPONENT}
%
%\end{align}
%
\noindent{where $p_{i0}$ yields the transmit power, $\alpha$ is the path-loss
exponent and $x_{i0}$ represents the composite fading channel.}
In what follows, when there is no fear of ambiguity let us omit the subscripts
of the interference component terms.
Fig. \ref{FIG:HN_GRID} illustrates a realization of the random network topology
where \acp{UE} and small cell \acp{BS} are uniformly scattered over network
area.

%% \reversemarginpar
%% \marginnote{R$2$C$1$}[0cm]
\section{Network deployment and Stochastic Geometry}
\label{SEC:SG_MODEL}

Active nodes constitute a homogeneous \ac{PPP} $\Phi$ with intensity $\lambda$
in $\mathbb{R}^2$.
The number of nodes in an arbitrary region $\mathcal{R}$ of area $A$ is a
Poisson \ac{RV} with parameter $\lambda A$
\cite{BOOK:KINGMAN-OXFORD93}.

%% \reversemarginpar
%% \marginnote{R$1$C$2$}[-2ex]
The fading effect is assumed as a random mark associated with each point of
$\displaystyle {\Phi}$, thus $\widetilde{\Phi} = \left\{ \left( \varphi, x
\right); \varphi \in \Phi \right\}$ corresponds to a {marked point process} on
the product space $\mathbb{R}^2 \times \mathbb{R}^+$, whose random points
$\varphi$ denoting transmitters locations and belong to $\Phi$
\cite{BOOK:BADDELEY-SPRINGER03, BOOK:KINGMAN-OXFORD93}.
The network deployment model is given by a spatial~\acp{PPP} $\Phi^\mathsf{BS}$
($\Phi^\mathsf{UE}$), whose random points $\varphi$ represent the locations of
\acp{BS} (\acp{UE}).
In addition, we consider a fixed number of potential relays that detect
source's message and take part in the selection procedure, it is well known
that such conditional distribution of points within an arbitrary forwarding
region $A$ is described by a binomial point process
\cite{BOOK:BADDELEY-SPRINGER03}.
As shown next, by employing stochastic geometry concepts to this deployment
model, it is possible to conveniently derive closed form expressions to
characterize the distributions of the received power and corresponding
aggregate \ac{CCI} with respect to the receiver of interest.
{Indeed, the path loss attenuation (function of the distance to the tagged user), the lognormal shadowing and Nakagami-$m$
    fading are combined using the framework of \cite{ART:GHASEMI-JSAC08,
    ART:LIMA-TWC12A} to obtain the distribution of the received power at the
user of interest.}

\subsection{Received Power from a Random Transmitter}
\label{SEC:PRX}

We resort to the framework introduced in \cite{ART:LIMA-TVT12B,
ART:LIMA-TWC12A} to attain the {characteristic function} and the respective
cumulants ($\kappa$) of the received power and aggregate interference of random
transmitters within the forwarding region of the user of interest.
Consider a \ac{RV} $Y = R^{-\alpha} X$ describing the power received
at the tagged receiver from a random transmitter in $\mathcal{O}$ (as defined
in Section~\ref{SEC:SYSTEM_MODEL}), with $R$ varying from $R_m$ to $R_M$,
and $X \thicksim \mathsf{Lognormal}(\mu_\mathrm{dB}, \sigma_\mathrm{dB})$
representing the squared-envelope of the composite fading channel.
Then, the {characteristic function} of $Y$ is \cite[Appendix
B]{ART:LIMA-TVT12B} $\Psi_Y \left ( \omega \right ) = {2} {(R_M^2-R_m^2)^{-1}}
\mathrm{E}_X\left [ \operatorname{R} \left( \omega \right) \right ]$ where
$\operatorname{R} \left( \omega \right) = \int_{R_m}^{R_M} { \exp{\left( j
\omega p r^{-\alpha} x \right)} r \mathrm{d}r }$ and $\mathrm{E}_X\left [
\,\cdot\, \right ]$ yields the expectation of the enclosed expression over
the \ac{RV} $X$, and is a general formulation which characterizes
the distribution of any random transmitter within the reception range of a
tagged receiver.
Thereafter, the $n\text{th}$ cumulant of $Y$ is given by \cite[Appendix
C]{ART:LIMA-TVT12B}
\begin{equation}
    \kappa_n = \frac{1}{\imath^n} \sum_{k=0}^n g^{\left( k \right)} \left( \beta_0
    \right) \negthinspace\cdot\negthinspace B_{n,k}\left[ \beta_1, \beta_2,
    \dots, \beta_{(n-k+1)} \right],
    \label{EQ:INTERFERENCE_COMPONENT_CUMULANT}
\end{equation}
{\noindent where $\imath$ yields the imaginary unity, $\displaystyle g \left (
    u \right ) = \ln {\left ( u \right ) }$, $B_{n,k}\left[ \beta_1, \beta_2,
    \dots, \beta_{(n - k + 1)} \right]$ is the partial Bell polynomial
    \textup{\cite{ART:BELL-TAM34}} and $\displaystyle\beta_n = \imath^n p^n
    \times \frac{R_m^{2-n \alpha } - R_M^{2-n \alpha }}{n \alpha - 2}
    \operatorname{E}_X \left[ x^n \right]$.} \\[1mm]

Next, we use the aforesaid cumulant-based framework to characterize the
aggregate interference at the user of interest for both the \ac{HD} and \ac{FD}
configuration scenarios.
Under the assumptions of Section \ref{SEC:SYSTEM_MODEL} and with respect to the
tagged receiver, we now use our analytical framework to characterize the
resulting aggregate \ac{CCI} in two \acp{ES}: \ac{ES}$1$ is the benchmark
scenario where \acp{BS} and \acp{UE} operate in \ac{HD} mode; while in
\ac{ES}$2$ both \acp{BS} and \acp{UE} operate in \ac{FD} mode.

\subsection{Aggregate \ac{CCI} from a Poisson Field of Interferers}
\label{SEC:PBS_CCI}

In our benchmark scenario, denoted \textit{\ac{HD} relaying}, all nodes operate
in \ac{HD} mode.
The corresponding aggregate \ac{CCI} at a tagged receiver is then given by
$Z_0^\mathsf{HD} =  \sum_{ \left( \varphi_i, x_i \right) \in \widetilde{\Phi}^\mathsf{BS} } Y_{i0}$.
%
%
%\begin{align}
    %
    %Z_0^\mathsf{HD} =  \sum\limits_{ \left( \varphi_i, x_i \right) \in \widetilde{\Phi}^\mathsf{BS} } \negthickspace Y_{i0}.
    %
    %\label{EQ:AGGREGATE_CCI_HD}
%
%\end{align}
%
Notice that the tagged receiver only experiences interference coming from
nearby {base stations} since we assume the \ac{DL} as the benchmark scenario.

In this context the aggregate \ac{CCI} at the tagged receiver is \cite[Section
V]{ART:LIMA-TWC12A}
\begin{equation}
    \kappa_n \negthickspace \left( \widetilde{\Phi}^\mathsf{BS} \right)
    =\frac{2 \pi \lambda\,p^n}{n \alpha - 2} \left( R_m^{2 - \alpha n}
    \negthickspace - R_M^{2 - \alpha n} \right) \operatorname{E}_X^n
    \negmedspace \left [ 0, \infty \right ].
    \label{EQ:CUMULANT_BS_FULL_INTERFERENCE}
\end{equation}

\subsection{Aggregate \ac{CCI} in the \ac{FD} Configuration}

%% \normalmarginpar
%% \marginnote{R$1$C$4$}[-2ex]
In the \ac{FD} configuration, the tagged receiver is subject to the
interference from \ac{FD} serving base stations, other \ac{FD} \acp{UE} and its
intrinsic \ac{SI} component.
{The self-interference channel between transmitting and receiving
    antennas of a full duplex transceiver exhibits a strong line of sight
    component and can be represented using a Ricean distribution with large
    $K$-factor \cite{ART:DUARTE-TWC2012} (a $K$-factor of $14.8$dB is used,
    which corresponds to a Nakagami distribution with the parameter $m =
16$).}
As a result, the aggregate \ac{CCI} at the tagged receiver is characterized
by
\begin{equation}
    Z_0^\mathsf{FD} = \delta p_{00} x_{00} + \sum\limits_{ \left( \varphi_i,
    x_i \right) \in \widetilde{\Phi}^\mathsf{BS} } \negthickspace Y_{i0} +
    \sum\limits_{ \left( \varphi_j, x_j \right) \in
    \widetilde{\Phi}^\mathsf{UE} } \negthickspace Y_{j0},
    \label{EQ:AGGREGATE_CCI_FD}
\end{equation}
{\noindent where $p_{00}$ and $\delta$ represent the \ac{SI}
component and the corresponding attenuation factor, respectively.}

Since \acp{BS} and \acp{UE} are assumed to be independently scattered over the
network deployment area, the resulting processes ($\Phi^\mathsf{BS}$ and
$\Phi^\mathsf{UE}$, respectively) from each such tier are also independent
\cite{BOOK:KINGMAN-OXFORD93}. 
Therefore, the cumulants additivity property is employed to obtain the
aggregate \ac{CCI} from multiple interfering tiers \cite{ART:LIMA-TWC12A}.
In other words, if $U$ and $V$ are independent \acp{RV}, then, we can
derive the cumulant of their sum as $\kappa_n (U + V) = \kappa_n (U) + \kappa_n (V)$.
%
%\begin{align}
%    %
%    \kappa_n (U + V) = \kappa_n (U) + \kappa_n (V).
%    %
%    \label{eq:cumulant_sum_independent}
%%
%\end{align}

\subsection{\acl{SIR} and Outage Probability}

By using the results from previous sections, it is straightforward to compute
the \ac{SIR} and outage probability of the evaluation scenarios under study.
The \ac{SIR} and outage probability are used to assess how the relay selection
strategies perform in \ac{HD} and \ac{FD} configuration mode as follows.
%
% The outage probability can be characterized as follows.
%
\begin{proposition}
    Let $V_{0}$ and $V$ be Normal \acp{RV} (in logarithmic scale)
    representing the power received from the desired transmitter and the
    \ac{CCI} at the receiver of interest.
    Under the assumption of the composite fading with Gamma-{lognormal}
    distribution, the \ac{SIR} at the tagged receiver is 
    $\Gamma \thicksim \mathsf{Normal} \left( \mu_{V_0}-\mu_V \,,\,
    \sigma^2_{V_0}   + \sigma^2_V \right)$,
    and the outage probability is given by 
    $\Pr\left [ \Gamma < \gamma_\text{th} \right ] = \operatorname{Q}\left[
    \left( \mu_{\Gamma} - \gamma_\text{th} \right)/\sigma_\Gamma \right]$,
    where $\mu_\Gamma = \mu_{V_0} - \mu_V$ and $\sigma^2_\Gamma =
    \sigma^2_{V_0} + \sigma^2_V$.
    Notice that the lognormal parameters can be attained as follows 
    $\mu = \ln{\left({\kappa_1^2} / {\sqrt{\kappa_1^2 + \kappa_2}}\right)},
    \thickspace \text{and} \thickspace \sigma^2 = \ln{\left( 1 + {\kappa_2} /
    {\kappa_1^2} \right)}$,
    where $\kappa_1$ and $\kappa_2$ are given by
    \eqref{EQ:INTERFERENCE_COMPONENT_CUMULANT}.
    \label{PROP:SIR_CORRELATED}
\end{proposition}
\begin{IEEEproof}
    The \ac{SIR} distribution is given by the quotient of two independent
    {lognormal} \acp{RV}, namely, $e^{V_0}$ which is the received
    power from the target transmitter, and $e^{V}$ which is an equivalent
    {lognormal} \ac{RV} approximating the aggregate \ac{CCI} at the
    tagged receiver \cite{BOOK:DALE-WILEY79}.
\end{IEEEproof}

\section{{\acl{RSA} and \acl{PGF}}}
\label{SEC:RSA}

Hereafter, we characterize the cost of selecting a suitable relay at hop basis
on the achievable throughput.
To do that, we first account for the relay selection overhead using the
\ac{PGF} of the \ac{CRI} and later combine the impact of the network
characteristics with the relay selection mechanism throughout a semi-Markov
process.
In this work, a totally random approach based solely on the {standard tree
algorithm} is used to implement the {relay selection algorithm}
\cite{ART:MATHYS-IT85}.
It is worth noticing we use the conditional \ac{PGF} of the \ac{CRI} length
($L_N$) that is derived in \cite{PROC:LIMA-ITW09,ART:LIMA-SJ2016} for the same
{standard tree algorithm} under investigation.
%
%Here, due to space limitations, we only summarize the main results regarding the relay selection procedure in the following section.
Next, due to space limitations, we only summarize the main results regarding
the relay selection procedure.% in the following section.

\subsection{{Relay Selection Algorithm}}

%% \reversemarginpar
%% \marginnote{R$2$C$2$}[-2ex]
Two relay selection approaches are considered in this work: ($i$) a static one
whereby a relay node is preassigned to forward source packets; and ($ii$) a
reactive strategy by which a relay node is identified based on the selection
criteria -- \textit{i.e.}, the neighboring node that provides the longest
advancement towards the final destination at each hop.
{While the former configuration may be more applicable to legacy
    networks where a smallcell may provide the last hop to cell edge users, the
    latter is tailored to upcoming $5$G networks where humans and machines
    communicate in large scale deployment scenarios
    \cite{PROC:POLESE-I3EICC16}.
    On one hand, the static relay procedure relies on long term characteristics
    of the deployment scenarios, \textit{i.e.} the relay is selected a priori
    following a network planning (it is arbitrary in our investigations) and
    does not change.
    On the other hand, the reactive relay procedure selects a random (new)
    relay within the transmission range every time a packet needs to be
    forwarded to the destination \cite{PROC:LIMA-ITW09,
    PROC:MARCHENKO-VTC11, ART:MARCHENKO-I3ETVT14}.}

Succinctly, the reactive relay selection algorithm is implemented through a
carrier sense collision avoidance mechanism similar to the IEEE $802.11$
handshake using \ac{RTS}/\ac{CTS} messages.
The source transmits a \ac{RTS} packet to initiate the relay selection
procedure.
Nodes that listen to this request reply with a \ac{CTS} packet based on the
predetermined probability of accessing the channel, $p$.
If a collision occurs, nodes that have transmitted in previous slot retransmit
or not based on a random access process similar to a $Q$-sided coin.
To select the best suitable relay (greed criteria), \textit{i.e.} the closest
node to the destination whether there is one available, the source node should
receive the replies from all the candidate relays
\cite{PROC:LIMA-ITW09, ART:LIMA-SJ2016}.
We consider the conditional \ac{CRI} length when $N$ nodes initially collide
(which corresponds to a {binomial point process}
\cite{BOOK:BADDELEY-SPRINGER03}).
Here, aiming at shorten the selection procedure\footnote{At the cost of
selecting a relay that is not necessarily the closest one to the final
destination}, the first node (at a random location within transmission range)
to successfully reply to the source is selected as the next hop relay -- note
that this mechanism differs from the one used in
\cite{PROC:LIMA-ITW09, ART:LIMA-SJ2016} which waits for the replies of all
neighboring relay candidates in order to select the most suitable based on
the predefined selection criteria.
\begin{figure}[!t]
	%	%%\vspace{-5mm}	
	\centering
	\begin{tikzpicture}[
	scale = 2,
	active/.style = {fill = black, line width = 1pt, circular drop shadow},
	idle/.style = {fill = white, line width = 1pt, circular drop shadow}
	]
	% Draws grid lines.
	\draw[step = 5mm, help lines, opacity = .5] (-2, -2) grid (2, 2);
	\fill (-2, -2) circle (2pt);
	%
	% Coordinates system.
	\draw[<->, line width = 1pt] (-1, -2) node[below] {$x$} -| (-2, -1)
	node[left] {$y$};
	%
	% Axes dashed lines.
	\draw[dashed, gray] (0, -2.5) -- (0, 2.5);
	\draw[dashed, gray] (-2.5, 0) -- (2.5, 0);
	%
	% Source and sink nodes.
	\node[left = .25cm] (src) at (0, 0) {\footnotesize src.};
	\filldraw[active] (-.075, -.075) rectangle +(.15,.15);
	\node[right = .25cm] (dest) at (3, 0) {\footnotesize dest.};
	\filldraw[active] (2.925, -0.075) rectangle +(.15,.15);
	%
	% Source's radio range.
	\clip[draw] (0, 0) circle (1.75cm);
	\filldraw[fill = none, line width = 1pt] (0, 0) circle (1.75cm);
	%
	% Decision region.
	\draw[dashed, line width = 1pt, fill = gray, fill opacity = .25] (0,0)
	-- (0,2) -- (1.75,2) -- (1.75,-2) -- (0,-2) -- cycle;
	\draw[dashed, line width=1pt, fill=darkblue, fill opacity=0.25] (0,0)
	-- (1.75,2) -- (1.75,-2) -- cycle;
	%
	% Relay nodes.
	\node[below right = .15cm] (a) at (.75, 1.35) {\footnotesize $21$};
	\filldraw[active] (.75, 1.35) circle (.1cm);
	\node[above left = .15cm] (b) at (1.5, -.25) {\footnotesize $53$};
	\filldraw[idle] (1.5, -.25) circle (.1cm);
	\node[below right = .15cm] (c) at (1, -.75) {\footnotesize $87$};
	\filldraw[active] (1, -.75) circle (.1cm);
	\node[above right = .15cm] (d) at (.5, -1.25) {\footnotesize $3$};
	\filldraw[idle] (.5, -1.25) circle (.1cm);
	\node[above right = .15cm] (e) at (.25, -.5) {\footnotesize $37$};
	\filldraw[idle] (.25, -.5) circle (.1cm);
	\node[above = .175cm] (f) at (.3, .15) {\footnotesize $55$};
	\filldraw[active] (.3, .15) circle (.1cm);
	\node[above right = .15cm] (g) at (-1, 1) {\footnotesize $1$};
	\filldraw[idle] (-1, 1) circle (.1cm);
	\node[above right = .15cm] (h) at (-1.5, -.5) {\footnotesize $11$};
	\filldraw[active] (-1.5, -.5) circle (.1cm);
	%
	% Transmission range and varying radius.
	\filldraw[fill = none, line width = 1pt, dashed] (0,0) circle(1cm);
	\node (a) at (-.35,.5) {\scriptsize $d$};
	\draw[->, line width=.6pt] (0,0) +(0:0cm) -- +(135:1cm);
	% \node (a) at (-1.25,.25) {\scriptsize $R$};
	% \draw[->, line width=.6pt] (0,0) +(0:0cm) -- +(155:1.75cm);
	%
	\end{tikzpicture}
	%\vspace{-3mm}
	\caption{Illustration of the sectoral decision region with $Q=2$ splitting
		groups and angular aperture of $180$ degrees.
		\textcolor{black}{Dashed lines defined the forwarding region (shaded
			regions in light gray and blue), black circles identify awake nodes and
			white circles identify asleep nodes.}
		All awake neighbors within the shaded region are eligible relays.}
	\label{FIG:CONFIG_SDR}
	%\vspace{-3mm}
\end{figure}
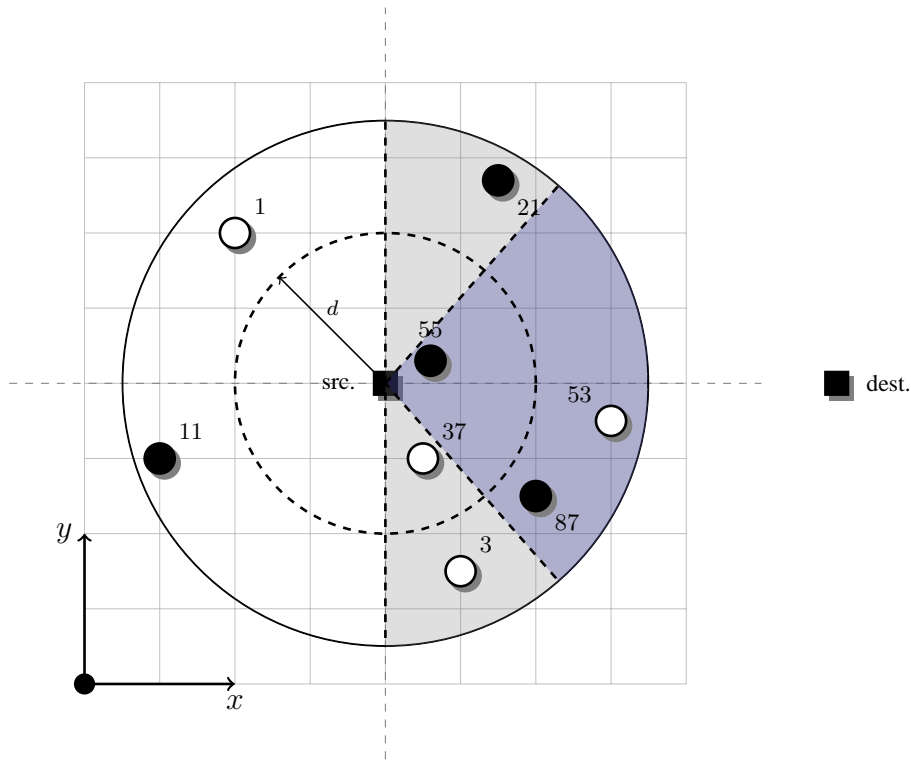

We use the relay selection mechanism introduced in
\cite{PROC:LIMA-ITW09,ART:LIMA-SJ2016} which combines the standard tree
algorithm to resolve contentions and sectoral decision regions to identify the
potential relays.
Fig. \ref{FIG:CONFIG_SDR} illustrates a snapshot of the sectoral decision
region that is used in conjunction with the standard tree-based solution.
In the next section, the \ac{PGF} is used to characterize the distribution of
the \ac{CRI} associated with the selection of a random relay at hop basis.
%
%% \normalmarginpar
%% \marginnote{R$1$C$6$}[-2ex]
{\begin{remark}
    Notice that current $802.11$ standards implements the RTS/CTS handshake as
    an optional feature to avoid frames collisions; besides, more recent
    standards such as $802.22$ and $802.11$af also incorporate cognitive radio
    functionalities which makes it even easier to deal with side-information
    \cite{PROC:LIMA-ITW09}.
\end{remark}}

{\begin{remark}
    In LTE systems, the contention-based random channel is based on ALOHA-type
    communication and is susceptible to performance degradation due to a high
    collision probability in the transmission of the preambles in dense
    deployments.
    Indeed, the contention-based operation of the LTE Physical Random Access
    Channel is also established by the signaling exchange of messages, namely,
    preamble transmission, random access response, connection request and
    contention resolution \cite{ART:LAYA-I3ECST14, PROC:POLESE-I3EICC16}.
\end{remark}}

\subsection{{\ac{PGF} of the \acl{RSA}}}

The conditional \ac{CRI} length considering a $Q$-sided fair coin is
generalized as,
\begin{equation}
    L_N =
    \left\{
        \begin{array}{l@{,\,}l}
            1 & \text{ if } N = 0, 1; \\
            1 + \sum\limits_{j = 1}^Q { L_{I_j} } & \text{ if } N \geq 2,
        \end{array}
    \right.
    \label{EQ:CRI_LENGTH_COND_N}
\end{equation}
{\noindent where $I_j$ is a discrete \ac{RV} describing the number of
candidate relays that tossed the $j$ value of the $Q$-side coin.}
For a binary splitting tree, the \ac{PGF} of the time spent by a random node in
the relay selection procedure is give by the following proposition.

\begin{proposition}
    Under the assumptions of a binary {standard tree algorithm} (unbiased coin)
    and $N$ conflicting nodes, the \ac{PGF} of the \ac{CRI} for a tagged packet
    is
    \begin{align}\label{EQ:G_PGF}
        &{G}_{N + 1}(z) = \frac{1}{2} \left[ {G}_{N + 1}^{(0)} (z) + {G}_{N +
        1}^{(1)} (z) \right] \text{ with } \\
        \label{EQ:G_PGF_0}
        &{G}_{N + 1}^{(0)} (z) = z \sum_{k = 0}^N B_{N, k} G_{k + 1} (z)
        \text{ and } \\
        \label{EQ:G_PGF_1}
        &{G}_{N + 1}^{(1)} (z) = z \sum_{k = 0}^N B_{N, k} Q_{k} (z) G_{N - k + 1}
        (z),
    \end{align}
    \label{PROP:G_PGF}
\end{proposition}
{\noindent where ${G}_{1} = z$, ${G}_{N + 1}^{(0)} (z)$ accounts for the
\ac{CRI} when the tagged node joins the first subset (flipped $0$), ${G}_{N +
1}^{(1)} (z)$ describes the resolution interval when the tagged node joins the
second subset (flipped $1$), and $B_{N, k}$ $= \Pr [k$ flipped $0$ $| N$
contending nodes$]$.
Note that the tagged packet may be in  each subset with equal probability.}
\begin{IEEEproof}
    Let $N > 1$ be the number of contending nodes that start the \ac{CRI}.
    For a fair binary splinting tree we have $B_{N, k} = \binom{N}{k} 2^{-k}$.
    Thus, if the first splitting group has $k$ members, the second contains the
    remaining $N - k$ nodes.
    Each such sub-\ac{CRI} is statically indistinguishable from a general
    \ac{CRI} with equal number of elements.
    Therefore, the recursive formula $Q_{N} (z) = \sum_{k = 0}^N B_{N, k} Q_{k}
    (z) Q_{n - k} (z)$ is obtained.
    By the same line of reasoning, we can defined 
    \begin{align}
        \label{eq:proof_cri_0}
        {G}_{N + 1} (z) &= \Pr [t_0 = k | N \text{ other nodes collide}] z^k;  \\
        \label{eq:proof_cri}
        {G}_{N + 1}^{(s)} (z) &= \Pr [t_0 = k | N \text{ other nodes collide}  \nonumber \\
        & \qquad \qquad \text{ and the tagged node flips } s] z^k,
    \end{align}
    {\noindent where $t_0$ yields the time a tagged packet spends in the
        \ac{CRI} before being successfully transmitted.}
    Finally, computing the corresponding probabilities for a fair binary
    spiting tree, we obtain \eqref{EQ:G_PGF}.
\end{IEEEproof}

In other words, Proposition \ref{PROP:G_PGF} gives the time a tagged packet
takes to be successfully transmitted.
When a random node within source's transmission range is selected, its distance
distribution is characterized by \eqref{EQ:INTERFERENCE_COMPONENT_CUMULANT}.
To determine the holding time matrix, we resort to the factorial moment concept
so as to compute the mean and variance of \ac{CRI} for a tagged packet from
Proposition \eqref{PROP:G_PGF}.
\begin{definition}
    Let $X$ be a discrete \ac{RV} taking non-negative integer values;
    then, the $k$th factorial moment of $X$ is given as  
%   % \begin{align} %
    $\operatorname{E}\left[ X (X - 1) \cdots (X - k + 1) \right] =
    G^{(k)}(1^-), \thinspace k \geq 0$ \cite{BOOK:HOWARD-DOVER07}.  
    %
%   \end{align}
    %
    \label{DEF:FACTORIAL_MOMENT}
\end{definition}
Then, we obtain the following results from Definition
\ref{DEF:FACTORIAL_MOMENT} %:
%
%\begin{align}
%    %
%    \label{EQ:FACTORIAL_MOMENTS}
%    G(1) &= \sum_k \Pr{[X = k]},  \\
%    %
%    \label{EQ:FACTORIAL_MOMENTS_1}
%    G^\prime (1) &= \operatorname{E} [X],  \\
%    %
%    \label{EQ:FACTORIAL_MOMENTS_2}
%    G^{\prime \prime} (1) &= \operatorname{E} [X (X - 1)].
%
%\end{align}
as $G(1) = \sum_k \Pr{[X = k]}$, $G^\prime (1) = \operatorname{E} [X]$ and 
$G^{\prime \prime} (1) = \operatorname{E} [X (X - 1)]$.
\section{{Semi-Markov Process}}
\label{SEC:SMP}
\begin{figure}[!t]
	\centering
	\includegraphics[width=.8\columnwidth]{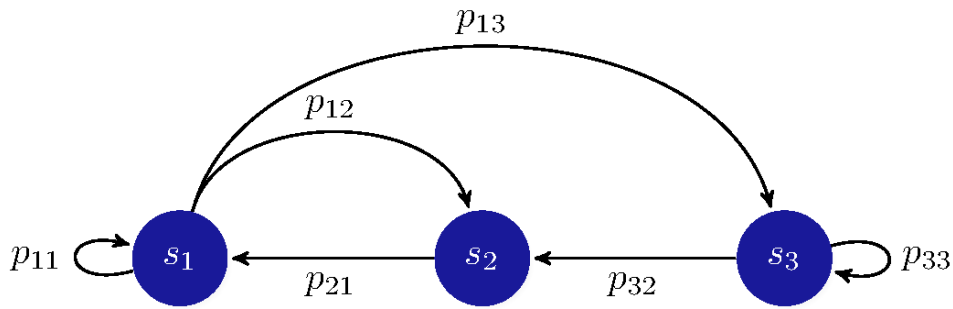}
	%\vspace{-3mm}
	\caption{Illustration of a Semi-Markov process with three states $s_i,~i
		\in \{1,2,3\}$.
		$s_1$ represents transmission, $s_2$ relay and $s_3$ retransmission.}
	\label{FIG:SMP}
	%
	%\vspace{-3mm}
\end{figure}

Markov processes permit represent random systems whose outcome at any given
instant depends only on the outcome that proceeds it (memoryless property, if
the present is specified, the past has no influence in the future).
A Markov chain is special kind of Markov process wherein the future evolution
of the process depends on the present state and not on how it arrived at that
state \cite{BOOK:PAPOULIS-MCGRAWHILL2002}.
Typically, such Markov models have the property that a transition occurs at
every instant -- even if the transition returns to the previous occupied state,
it occurs anyway.
In this work, we use a semi-Markov process that is associated with the embedded
Markov chain in Fig. \ref{FIG:SMP} that tracks the relay selection protocols
states.
The semi-Markov process theory allow us to obtain the steady-state average
delay and throughput during a cycle taking any state of the chain as a
reference \cite{BOOK:PAPOULIS-MCGRAWHILL2002, BOOK:HOWARD-DOVER07}.
Hence, at transition instants semi-Markov process behaves just as a Markov
process.
is used to combine the network topology (including radio channel) that are
captured by the stochastic geometry with the routing protocol dynamics that are
captured by the \ac{PGF} framework.
With semi-Markov process, the elapsed time between transitions may take several
unit intervals, while at transition instants it behaves as a typical Markov
process \cite{BOOK:HOWARD-DOVER07}.
The semi-Markov process depends on the transition that occurred, though the
transition instants follow a distinct probabilistic mechanism
\cite{BOOK:HOWARD-DOVER07}.
Throughout our investigation we consider the states presented in Fig.
\ref{FIG:SMP}: $s_1$, $s_2$ and  $s_3$ corresponds to the transmission, relay
and re-transmission states, respectively.
The corresponding state transition matrix is given by
\begin{equation}
\mathbf{P}= \begin{bmatrix}
        p_{SD}\! & (1 - p_{SD}) p_{SR}\! & (1 - p_{SD}) (1 - p_{SR}) \\
        1\! & 0\! & 0 \\
        p_{SD}\! & (1 - p_{SD}) p_{SR}\! & (1 - p_{SD}) (1 - p_{SR})
    \end{bmatrix}
    \label{EQ:TRANSITION_MTX}
\end{equation}
{\noindent where each element $p_{ij}$ yields the transition probability from
state $i$ to $j$, as well as $p_{SD}$ and $p_{SR}$ yield the probability that
the source message is correctly received by the final destination and relay
node, respectively.}

%% \reversemarginpar
%% \marginnote{R$3$C$3$}[-2ex]
{\begin{remark}
    The Markov process model has the property that a transition is made at
    every time instant.
    Differently, the semi Markov model is a more general class of processes
    where the elapsed time between transitions may take several unit time
    intervals.
    Then, it brings much more flexibility to the problem of modeling dynamic
    probabilistic systems.
    In fact, the successive state occupancies are governed by the transitions
    probabilities of an embedded Markov process, while the holding time in each
    state follows an integer-valued random distribution -- the conflict
    resolution interval, see \eqref{EQ:CRI_LENGTH_COND_N} in Section
    \ref{SEC:RSA} -- which depends on both the current and the next states
    \cite{BOOK:HOWARD-DOVER07}.
\end{remark}}

The \ac{PGF} of the \ac{CRI} $G_N (z)$ characterizes the relay selection
overhead at each hop.
The probability that a certain \ac{CRI} takes $k$ slots can be retrieved from
the respective \ac{PGF} using the following proposition.
\begin{proposition}\label{PROP:pmf}
    The probability mass function of $L_N$, the length of the contention
    resolution interval (in transmission slots) conditional on the number of
    contending nodes is $\Pr [X = x] = {G^{(x)}(0)}/{x!}$.
    %
%    \begin{align}
%        %
%        \Pr [X = x] = \frac{G^{(x)}(0)}{x!}.
%        %
%        \label{eq:pmf}
%    %
%    \end{align}
%
\end{proposition}

\begin{IEEEproof}
    From the \ac{PGF} definition it is straightforward that
    \begin{equation}
        G_N (z) = \operatorname{E} [z^X] %\nonumber \\        &
        = \sum_{x = 0}^\infty \Pr [X = x] z^x .
        \label{eq:pgf_def}
    \end{equation}
    Then, taking the $n$th derivative of \eqref{eq:pgf_def} and making $z = 0$ we attain the result in Proposition~\ref{PROP:pmf}.%we obtain \eqref{eq:pmf}.
\end{IEEEproof}

The holding time matrix incorporates into the embedded Markov chain of Fig.
\ref{FIG:SMP} the time spent in a given state between successive transitions.
\begin{align}
    \mathbf{H} = \begin{bmatrix}
        1 & 1 + L_N &  1 + L_N \\
        1 & 1 & 1 \\
        1 & 1 + L_N &  1 + L_N
    \end{bmatrix},
    \label{EQ:HOLDING_MTX}
\end{align}
{\noindent where each element $h_{ij}$ yields the time elapsed in state $i$
before transition to $j$.}

\subsection{Steady -- State Throughput Efficiency}

Let $i$ be an arbitrary state chosen as reference,  and $\pi_i$ be its
steady-state probability where the vector $\pi$ corresponds to the stead-state
distribution of the embedded Markov chain (see Fig.  \ref{FIG:SMP}).
The time delay associated to the transition from state $i$ to $j$ is given by
$D_{ij}$.
Similarly, $R_{ij}$ is the reward associated with the transition from state $i$
to $j$.
A cycle is defined as the time between two consecutive passages for the
reference state states.
A message is successfully delivered to the destination node when the process
returns to the state $s_1$.
The reward function $\operatorname{R} (\tau)$ corresponds to the cumulative
reward after a period $\tau$ and yields the total amount of correct
receptions.
According to the fundamental renewal-reward theorem \cite{BOOK:HOWARD-DOVER07},
the long-term throughput is given by the following value,
It is worth noticing that the resulting steady-state throughput efficiency is
independent of the choice of the reference state.
%
%% \normalmarginpar
%% \marginnote{R$1$C$2$}[-2ex]
{Moreover, the long term evaluation of $\displaystyle
    {\operatorname{R} (\tau)}/{\tau}$ actually corresponds to the normalized
    throughput of the protocol as introduced in \cite{ART:ZORZI-I3ETC96} and
    revisited in \cite{ART:MARCHENKO-I3ETVT14}}
\begin{equation}
    \eta = \lim_{\tau \to \infty} \frac{\operatorname{R} (\tau)}{\tau}
   % \nonumber \\ &
    %
    =\frac{\sum_{i = 1}^K {\pi_i \overline{\operatorname{R}}_i}}{\sum_{i = 1}^K
    {\pi_i \overline{\operatorname{D}}_i }},
    \label{eq:throughput}
\end{equation}
{\noindent where $K$ yields the number of states, $\overline{\operatorname{R}}
= \sum_j^K p_{ij} \operatorname{R}_{ij}$ is the mean rewarding, while
$\overline{\operatorname{H}} = \sum_j^K p_{ij} \operatorname{h}_{ij}$ is the
mean waiting time.}

\section{{Numerical Results}}
\label{SEC:NUMERICAL_RESULTS}

%% \reversemarginpar
%% \marginnote{R$1$C$8$}[-4ex]
For both desired link and interferers, a composite fading channel model with
Nakagami-$m$ parameter of $m=16$ (which corresponds to a Ricean fading channel
with factor $K = 14.8$dB) and shadowing standard deviation of $10 \thinspace
\text{dB}$ are considered as well as channel reciprocity between a
communication pairs.
Interferes are scattered over the network area with $R_m=25$ and $R_M=500$
meters, while $\lambda = 5 \times 10^{-5} \text{nodes}/\text{m}^2$ is
considered (which corresponds to nearly $40$ interferers in average) and the
path loss attenuation is set to $3$.
The \ac{CRI} is conditioned on the mean number of candidate relays on the
forward region of the source node.
Without loss of generality, we consider three contending relays within source's
forwarding region to compute the relay selection interval.
We assume that data and control information occupies one transmission interval.

% \subsection{Numerical Results}
%
\begin{figure}[!t]
    \centering
    \includegraphics[width=\columnwidth]{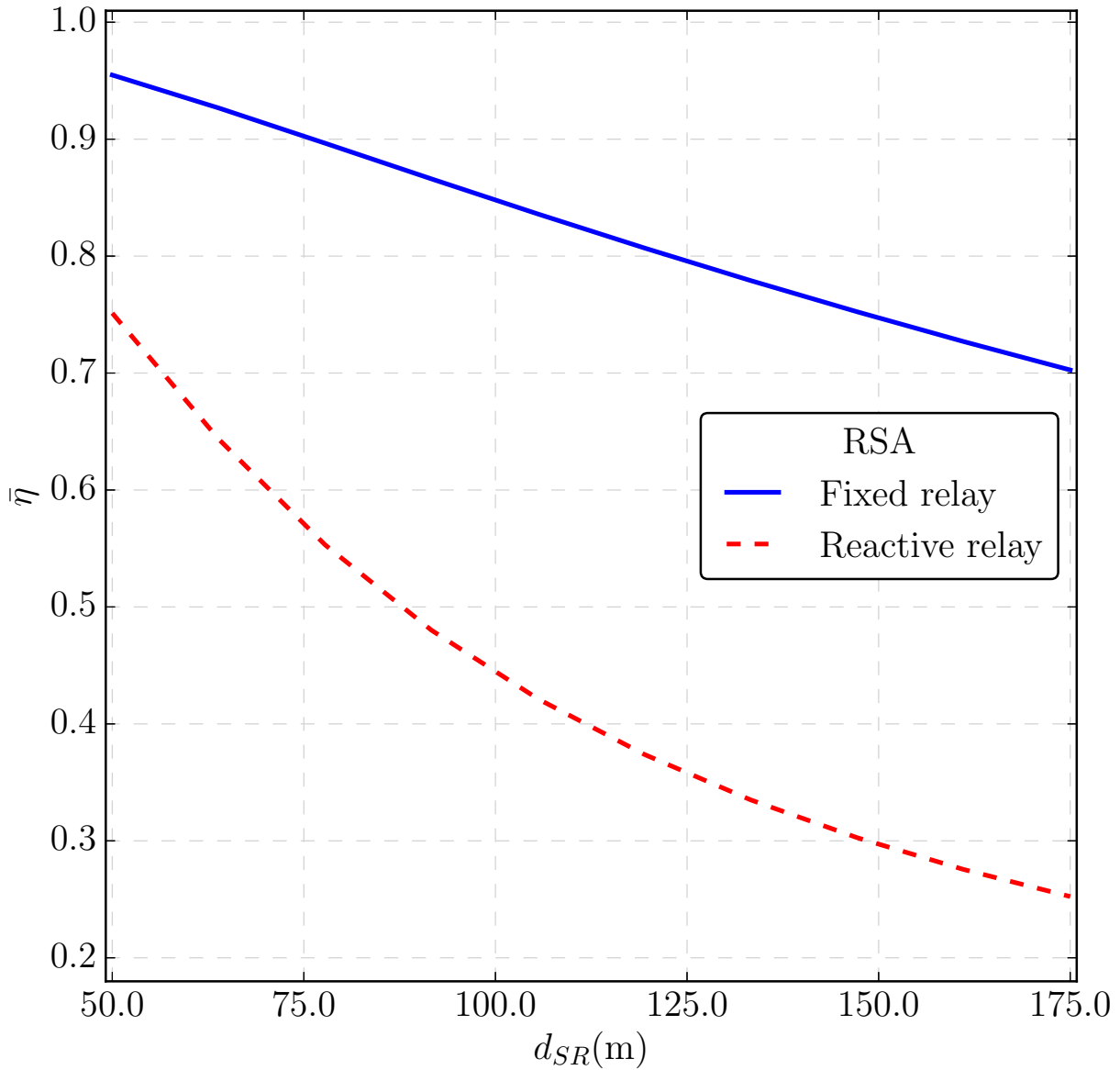}
    %\vspace{-5mm}
    \caption{Steady-state throughput efficiency for increasing separation
        distance between source -- destination pair in \ac{HD} configuration.}
    \label{FIG:HD_THROUGHPUT_DIST}
    %\vspace{-4mm}
\end{figure}
In a \ac{HD} configuration, Fig. \ref{FIG:HD_THROUGHPUT_DIST} shows the
steady-state throughput efficiency $\bar{\eta}$ for increasing separation
distance between source--destination pair.
\begin{figure}[!t]
    \centering
    \includegraphics[width=.9\columnwidth]{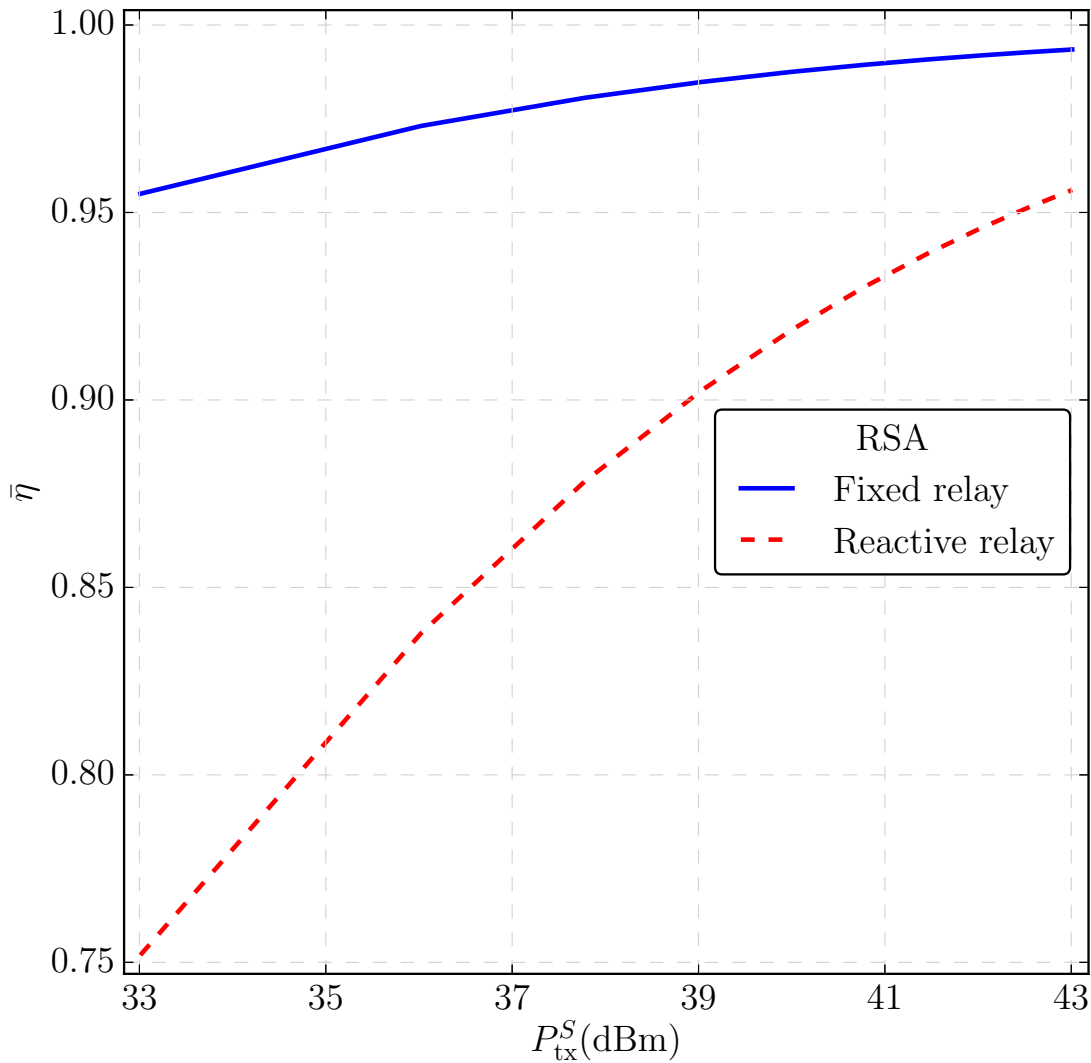}
    %\vspace{-5mm}
    \caption{Steady-state throughput efficiency increasing source's
        transmission power in \ac{HD} configuration.}
    \label{FIG:HD_THROUGHPUT_PTX}
    %\vspace{-4mm}
\end{figure}
As evidenced by the reactive relay curve (red dashed line), the  steady-state
throughput is severely compromised by the relay selection. 
On the other hand, when a fixed relay is considered within source's
transmission range, $\bar{\eta}$ becomes much less susceptible to the degrading
effect of longer separation distance between source and destination.

Fig. \ref{FIG:HD_THROUGHPUT_PTX} presents the steady-state throughput for
increasing transmission power of source node, whereas interferers transmit at
$30 \thinspace \text{dBm}$.
The separation distance between source and destination is kept at $50
\thinspace \text{m}$ and the relay is randomly located in the source's
forwarding region.
As expected the steady-state throughput improves with higher source's
transmission power.
The reactive relay benefits most from high transmission power mainly because
source can reach the destination more often without undergoing long relay
selection intervals to select a suitable relay.
\begin{figure}[!t]
    \centering
    \includegraphics[width=.9\columnwidth]{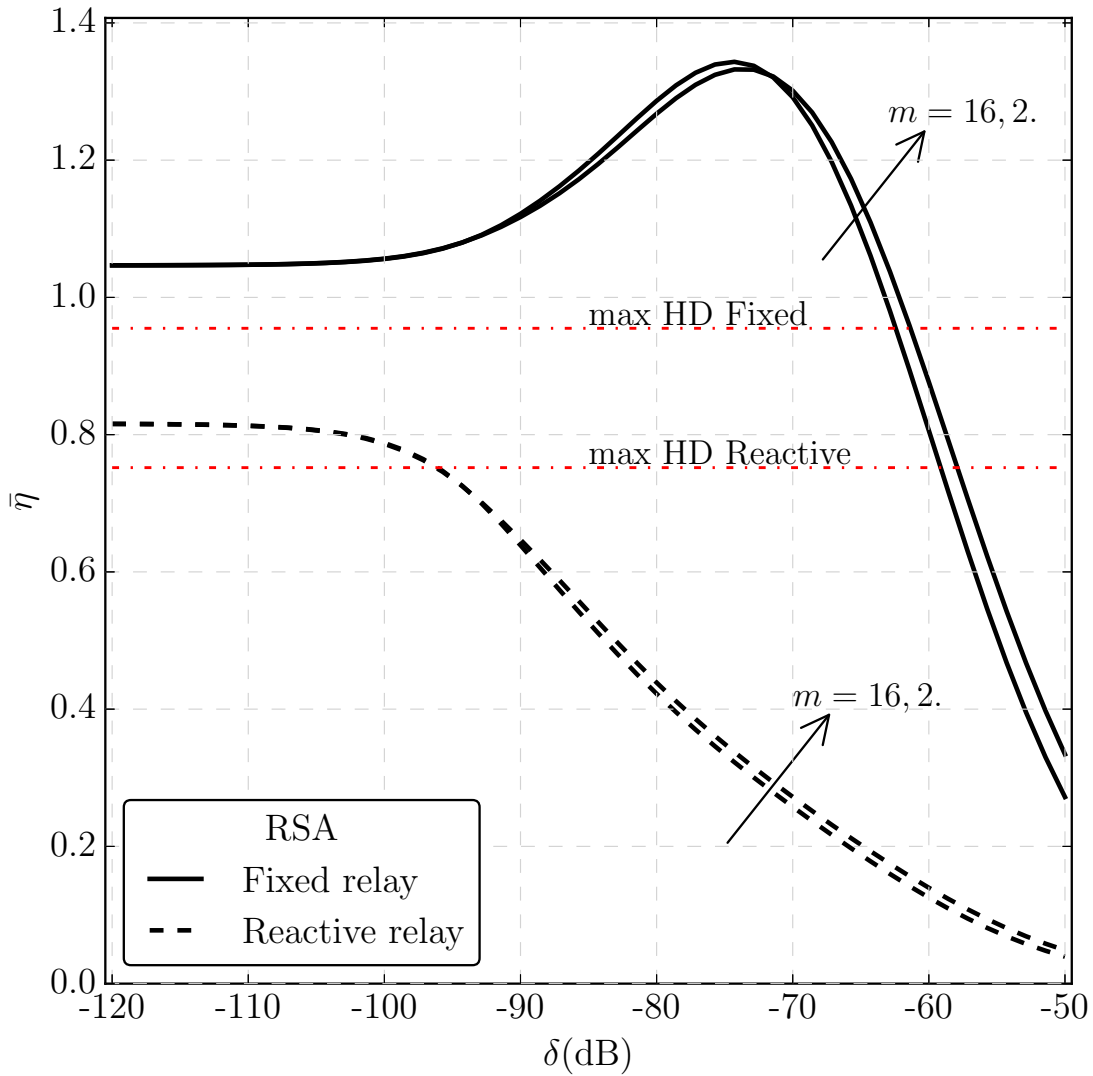}
%\reversemarginpar
%\marginnote{R$1$C$2$}[-4.25cm]
	%\vspace{-5mm}
    \caption{\hlchange{Steady-state throughput efficiency for decreasing SI attenuation
        in \ac{FD} configuration.}}
    \label{FIG:FD_THROUGHPUT}
    %\vspace{-4mm}
\end{figure}

The steady-state throughput efficiency for decreasing SI attenuation with
\ac{FD} configuration is shown in Fig. \ref{FIG:FD_THROUGHPUT}.
The separation distance between source--destination is set to $50 \thinspace
\text{m}$.
As previously observed in the \ac{HD} configuration, the steady-state
throughput with reactive relay is severely degraded by the relay selection
procedure and barely outperforms the \ac{HD} scenario with $100 \thinspace
\text{dB}$ SI attenuation.
\hlchange{It is also shown how the steady-state throughput efficiency scales
for $m = 2$ and $16$.
As can be seen, $\overline{\eta}$ slightly degrades when the Nakagami-$m$
parameter of the self-interference component increases from $2$ (non line of
sight) $16$ (line of sight).}

Regarding the fixed relay scheme, the \ac{FD} configuration with SI attenuation
raging from $100$ to nearly $70 \thinspace \text{dB}$ show much better
performance than \ac{HD} mode, whereas the performance degrades faster with
lower attenuation values.
From $120$ to $100 \thinspace \text{dB}$ the \ac{FD} gain is negligible,
because \ac{SI} is low and the source node reaches the destination directly.
However, for lower \ac{SI} attenuation the source node uses the relay more
often which increases $\bar{\eta}$ (the reward doubles).
%
%% \reversemarginpar
%% \marginnote{R$1$C$5$}[-.75cm]
Despite that, if the attenuation is lower than $70 \thinspace \text{dB}$, the
\ac{SI} also compromises the relay performance and degrades the Steady-state
throughput efficiency considerably.
Moreover, as indicated by the dot-dashed threshold lines, the FD configuration
outperforms the HD one for high \ac{SI} attenuation values.
However, if the \ac{SI} attenuation drops below $70 \thinspace \text{dB}$, the
FD performance degrades and HD becomes a better alternative.
In fact, it would be advantageous to consider a dynamic forwarding scheme by
which potential relays would change between FD and HD depending on the
attainable \ac{SI} value.

\section{Conclusions and Final Remarks}
\label{SEC:CONCLUSION}

In this work, we assess how in-band full duplex relaying performs for distinct
selection strategies, namely fixed and reactive.
Our investigations are carried out using an analytical framework based on
stochastic geometry and semi-Markov process.
The former is used to model the network deployment area and radio channel,
while the latter characterizes the cost of the relay selection procedure under
study.
Our results show that the benefits of the \ac{FD} configuration are compromised
by the frequent selection of the next hop relay.
Moreover, the steady--state throughput degrades depending on the relative
distance between source, relay and destination, as well as the respective
\ac{SI} attenuation.
%
%% \reversemarginpar
%% \marginnote{R$1$C$5$ R$1$C$9$}[-.75cm]
{As pointed out in \cite{PROC:LIMA-ITW09,ART:LIMA-SJ2016}, a binary search tree
implementation does not take full advantage of the side information (local
network topology) to shorten the relay selection time.
    Hence, we plan to extend this work by considering Dutch auction-based relay
    selection to exploit the local topology of the network in order to shorten
    the selection transactions.
    We also expect to assess the secrecy of the network, as in
    \cite{Hirley2015a}, by employing a similar framework.}
    %% %
    %% We also expect to assess the secrecy of the network, as in \cite{Hirley2015a, 
    %% Nardelli2015a}, by employing a similar framework.

\footnotesize
\bibliographystyle{IEEEtran}
\footnotesize
\bibliography{IEEEabrv,fd_markov}

\end{document}